# Tuning thermo-magnetic properties of dilute-ferromagnet multilayers using RKKY interaction

D. M. Polishchuk\*, M. Persson, M. M. Kulyk, E. Holmgren, G. Pasquale, and V. Korenivski\*

*Nanostructure Physics, Royal Institute of Technology, Stockholm 10691, Sweden*

E-mail: dpol@kth.se, vk@kth.se

We demonstrate a 20-fold enhancement in the strength of the RKKY interlayer exchange in dilute-ferromagnet/normal-metal multilayers by incorporating ultrathin Fe layers at the interfaces. Additionally, the resulting increase in the interface magnetic polarization profoundly affects the finite-size effects, sharpening the Curie transition of the multilayer, while allowing to separately tune its Curie temperature via intralayer magnetic dilution. These results should be useful for designing functional materials for applications in magneto-caloric micro-refrigeration and thermally-assisted spin-electronics.

Synthetic antiferromagnets (SAFs), fabricated either as continuous films or arrays of nanopillars, are magnetic multilayers composed of alternating ferromagnetic layers and nonmagnetic spacers, in which the layers' magnetic moments are arranged in antiparallel. Since the discoveries of the oscillatory interlayer exchange coupling[1] and giant magnetoresistance,[2,3] SAFs have enabled a range of spintronic devices[4,5] such as spin-valve sensors[6,7] and magnetic random access memory.[8–10] SAFs favorably combine the advantages of both ferromagnetic and antiferromagnetic materials. For example, nanopatterned SAFs have negligible magnetic stray fields akin to antiferromagnets, which reduces crosstalk problems in magnetic memory arrays. At the same time, non-zero magnetization of the individual ferromagnetic layers comprising the SAF can be used to control its configuration by external magnetic fields as well as sense its individual stable magnetic states using spin currents – the tasks hardly achievable with ordinary antiferromagnets. Continuous SAF multilayers usually incorporate rather strong interlayer exchange known as the Ruderman-Kittel-Kasuya-Yosida (RKKY) interaction.[11,12] RKKY-coupled SAF materials demonstrate significant potential for such promising devices[13] as the race-track memory[14,15] and the skyrmion-based logic.[16,17]

The useful functionality of SAF-systems extends to the realm of thermo-magnetic effects.[18–23] Incorporating dilute-ferromagnets (e.g., $Fe_xCr_{1-x}$) with relatively low Curie point ($T_C$ near room temperature) into RKKY-coupled Fe/Cr-based multilayers was experimentally shown to enable thermally-controlled antiferromagnetic exchange[21,22] as well as a giant magnetocaloric effect.[23] These were explained in terms of a thermally-driven competition between the intra- and inter-layer exchange interactions, when the two are tuned to be comparable in magnitude. Since the interlayer RKKY-exchange is an interfacial effect and is usually much weaker that the intra-layer exchange, an addition of an ultra-thin strongly-ferromagnetic layer (e.g., Fe) at the $Fe_xCr_{1-x}$/Cr interfaces allows to substantially enhance the interlayer coupling. The addition of Co layers at the interfaces of Permalloy has been shown[24,25] to enhance the interlayer exchange by up to several fold. The same effect of the interface polarization enhancement in lower-$T_C$ dilute-ferromagnet-based SAFs has not been studied in detail.

In this Letter, we demonstrate a 20-fold increase in the strength of the interlayer RKKY coupling in low-$T_C$ $[Fe_{37}Cr_{63}/Cr]_N$ multilayers by incorporating 0.25-nm-thick Fe layers at the Fe-Cr/Cr interfaces. The multilayers being either exchange coupled ($d_{Cr}$ = 1.2 nm) or decoupled ($d_{Cr}$ = 3.0 nm) were studied using vibrating-sample magnetometry at room temperature while varying the thickness of the $Fe_{37}Cr_{63}$ layers. The obtained values of the saturation magnetization and the interlayer exchange constant reveal profoundly different thickness dependence for the multilayers with and without interfacial Fe layers. The results are explained in terms of a complex interplay between the thermally-enhanced finite-size effects and the Fe-enhanced magnetic polarization of the interfaces, well supported by atomistic spin simulations performed using the VAMPIRE software package.[26]

Multilayers [Fe$_{37}$Cr$_{63}$($d_f$)/Cr($d_{Cr}$)]$_N$ (hereafter Fe-Cr/Cr) were grown on undoped Si (100) substrates at room temperature using an UHV dc magnetron sputtering system (AJA Inc.). Layers of a dilute Fe$_{37}$Cr$_{63}$ binary alloy were deposited using co-sputtering from separate Fe and Cr targets. The Cr thickness of $d_{Cr}$ = 1.2 nm was found to correspond to the strongest antiferromagnetic interlayer exchange coupling. At the same time, $d_{Cr} \geq$ 3.0 nm was determined to correspond to a completely decoupled state of the Fe-Cr layers. Since the interface roughness increases with the number of bilayer repetitions and can affect the RKKY interaction of the top layers, we have found the optimal number of the Fe-Cr/Cr bilayers to be $N$ = 8, for which this effect is negligible.

Two characteristic sample series are studied in detail – the base multilayered structure [Fe$_{37}$Cr$_{63}$/Cr]$_8$ (series Fe-Cr/Cr) and the one with about a monolayer (nominal thickness of 0.25-nm) of pure Fe added at the Fe-Cr/Cr interfaces (series Fe-Cr/Fe/Cr). Each sample series includes RKKY-coupled ($d_{Cr}$ = 1.2 nm) and decoupled ($d_{Cr}$ = 3.0 nm) configurations, with the Fe-Cr layer thickness varied as $d_f$ = 1, 2, 3, and 5 nm. The bulk Fe$_{37}$Cr$_{63}$ alloy shows the Curie temperature of ~370 K, found to be optimal for illustrating the physics involved using room temperature vibrating-sample magnetometry (VSM, Lakeshore Inc.), with the magnetic field applied in the film plane.

Figure 1(b) shows *M-H* curves for Fe-Cr(5 nm)/Cr structures with ($d_{Cr}$ = 1.2 nm) and without ($d_{Cr}$ = 3.0 nm) interlayer RKKY coupling. The zero remanence and the relatively high saturation field of the *M-H* curve ($d_{Cr}$ = 1.2 nm; orange) are the defining characteristics of an antiferromagnetically coupled SAF. The negligible coercivity and the low remanence of the black *M-H* curve ($d_{Cr}$ = 3.0 nm, with negligible RKKY exchange) indicate a highly thermally-agitated state of the thin Fe-Cr(5 nm) layers due to the proximity of the measurement temperature to the Curie point ($T_C^{bulk} \approx$ 370 K). The saturation field $H_s$, determined as the field-point where the two curves merge, reflects the strength of the interlayer RKKY coupling.

Decreasing the thickness of the Fe-Cr layers leads to a dramatic reduction in the saturation magnetization [Fig. 1(c)], indicating a strongly rising thermal spin disorder and finite-size effects.[27,28] For larger surface-to-volume ratios, the number of the stronger thermalized surface spins increases with respect to the stronger exchange-stabilized inner spins, resulting in lowering of the effective Curie temperature ($T_C^*$) for thinner ferromagnetic layers. One can observe that Fe-Cr/Cr with $d_f$ = 2 nm is paramagnetic, with $T_C^*$ lower than room temperature.

The interface-enhanced Fe-Cr/Fe/Cr structures exhibit *M-H* curves (Fig. 2) that are quite different from those for the structures without interfacial Fe [Fig. 1(c)]. First, they show an order of magnitude larger saturation fields for $d_{Cr}$ = 1.2 nm, which indicates a substantial increase in the strength of the interlayer RKKY coupling. Second, the structures remain ferromagnetic and show strong interlayer coupling for Fe-Cr as thin as 1 nm. Apparently, the interfacial Fe near-monolayer considerably modifies both the intra- and inter-layer exchange in the system.

The strength of the interlayer RKKY coupling is characterized by the interlayer exchange constant, $J_{ex} = 0.5 M_s H_s d_f$, where saturation magnetization $M_s$ and saturation field $H_s$ can be determined from the measured *M-H* curves. Figure 3 compares $M_s$ and $J_{ex}$ obtained for the RKKY-coupled structures. As seen from Fig. 3(b), the addition of interfacial Fe can increase $J_{ex}$ by more than 20 times. This is a very strong enhancement considering that the maximum increase of the interlayer coupling using this approach was about 4 times, reported for $Fe_{81}Ni_{19}$/Co/Ag multilayers.[25] As we show in greater details below, it can be explained in terms of a considerable difference between the effective interatomic magnetic exchange (magnetic polarization) and the accompanying thermal disorder for the diluted inner versus the surface Fe-rich spins. This interplay is reflected in the behavior of the saturation magnetization as follows.

$M_s$ of the Fe-enhanced structures (Fe-Cr/Fe/Cr) shows relatively small changes with varying the Fe-Cr thickness, which is in contrast to the steep rise in $M_s$-vs-$d_f$ for the base layout [Fe-Cr/Cr; Fig. 3(a)]. The addition of interfacial Fe has two relevant effects, both enhancing the magnetization: it stabilized the interface spins against thermal agitation, and it increases the overall atomic proportion of Fe to Cr in the structure. With increasing the thickness of the Fe-Cr layers (their inner section), the Fe-to-Cr ratio decreases while the finite size effects weaken, which explains almost constant $M_s$ vs. $d_f$.

The finite-size effects, thermal agitation, as well as interfacial alloying should be expected to result in pronounced changes in the thickness profile of the local magnetization in the ultra-thin dilute ferromagnetic layers used in this study. We have performed atomistic modeling of the interplay of these effects in our system using the VAMPIRE software package.[26]

Figure 4(a) shows the local magnetization versus the monolayer's depth, $m$-vs-$z$, for a 17-monolayer-thick $Fe_{37}Cr_{63}$-alloy film with and without a 2-ml-thick interfacial Fe layer; the simulation temperature chosen was $T/T_{C0} = 0.55$, where $T_{C0}$ is the Curie temperature of the bulk $Fe_{37}Cr_{63}$ alloy. The layer without interfacial Fe, plotted in orange in Fig. 4(a), exhibits much lower local magnetization at the interfaces, which supports the interpretation of the thermally-enhanced finite-size effects (interface spin disorder) discussed above. In contrast, the $m$-vs-$d_f$ profiles for the structures with interfacial Fe reveal significantly enhanced local magnetization at the interfaces, shown in blue in Fig. 4(a). The large difference in the interfacial magnetization for the two systems explains the observed large enhancement in the interlayer RKKY coupling, which should be proportional to the magnetic polarization of the interfaces.

The simulated total magnetization as a function of the Fe-Cr thickness, shown in Fig. 4(b), is in good agreement with the experimental data shown in Fig. 3(a). Additional simulations of the temperature dependence of the magnetization, presented in Fig. 4(c),(d), explain the thickness-dependent changes as due to changes in the respective Curie temperature, $T_C$. Predictably, with decreasing the Fe-Cr thickness, $T_C$ becomes higher or lower for Fe-Cr layers with or without interfacial Fe [Fig. 4(d)]. It is worth to note that the inner part of the model Fe/Fe-Cr/Fe structures show a sharper temperature dependence and lower values of $T_C$; cf. dashed curve in Fig. 4(c) This explains the rather sharp temperature transitions, as narrow as 15 K, observed for similar Fe-Cr based multilayers.[22] While the interlayer RKKY coupling (proportional to the interfacial magnetization enforced by locally adding pure Fe) can persist, the intra-layer exchange coupling can vanish at a given temperature determined by the amount of magnetic dilution in the Fe-Cr layers. The result is that one can separately tune the RKKY and $T_C$ in the structure.

The demonstrated strongly RKKY-coupled dilute-ferromagnet multilayers having the $T_C$ tunable in a wide temperature range can form the basis for designing functional materials with specific thermo-magnetic properties. An example could be improved multilayers showing ON/OFF switching of the antiferromagnetic interlayer exchange in the temperature interval as narrow as 15 K.[22] The low magnetization and relatively strong interlayer coupling in the close vicinity of $T_C$ can be interesting for thermally assisted spintronic applications.[29,30] On the other hand, the comparable strength of the intra- and inter-layer exchange and widely tunable $T_C$ can be promising for miniature magnetocalorics.[31–33]

In conclusion, the incorporation of ultra-thin Fe layers at the interfaces of dilute-ferromagnet [Fe-Cr/Cr]$_N$ SAFs yields an up to 20-fold increase in the strength of the antiferromagnetic interlayer coupling, explained as due to the Fe-enforced magnetic polarization of the interfaces diminishing the intertwined finite-size and thermal spin-disorder effects. The experimental findings are supported by numerical atomistic modeling, showing spin profiles of the local magnetization to be highly sensitive to the presence of additional interfacial Fe and magnetic dilution of the ferromagnetic layers. The demonstrated wide-range RKKY-vs-$T_C$ tunability should be interesting for spintronic and magneto-caloric applications.

**Acknowledgments**

Support from the Swedish Research Council (VR Grant No. 2018-03526), the Swedish Stiftelsen Olle Engkvist Byggmästare, and the Swedish Institute Visby Programme 2019/20 are gratefully acknowledged.

## Figure Captions

**Fig. 1.** (a) Multilayer layout of Fe-Cr/Cr SAFs and (b) their *M-H* curves, shown for typical cases with and without RKKY coupling (thin and thick Cr spacer, $d_{Cr}$ = 1.2 and 3.0 nm, respectively). (c) In-plane *M-H* curves for Fe$_{37}$Cr$_{63}$($d_f$)/Cr($d_{Cr}$), with $d_{Cr}$ = 1.2 and 3.0 nm, for different thicknesses $d_f$ of Fe-Cr layers, $d_f$ = 2, 3, 5 nm; vertical arrow indicates the saturation field $H_s$.

**Fig. 2.** *M-H* curves for Fe$_{37}$Cr$_{63}$($d_f$)/Fe(0.25 nm)/Cr ($d_{Cr}$ = 1.2, 3.0 nm) for Fe-Cr thickness $d_f$ of 1, 2, 3, and 5 nm.

**Fig. 3.** (a) Saturation magnetization, $M_s$, and (b) interlayer exchange constant, $J_{ex}$, for Fe$_{37}$Cr$_{63}$($d_f$)/Cr(1.2 nm) SAFs (with and without interfacial Fe) as a function of Fe-Cr layer thickness.

**Fig. 4.** Atomistic VAMPIRE simulations for a model thin-film system Fe(0/2 ml)/random-alloy-Fe$_{37}$Cr$_{63}$($d_f$)/Fe(0/2 ml). (a) Thickness profiles of local magnetization, $m_x$, for $d_f$ = 17 ml (~3 nm). (b) $d_f$ dependence of total magnetization, $<m_x>/m_0$; $m_0$ is zero-temperature magnetization. Simulations were performed at $T/T_{C0}$ = 0.55, where $T_{C0}$ is Curie temperature of bulk Fe$_{37}$Cr$_{63}$ alloy. (c) Temperature dependence of $<m_x>$ for $d_f$ = 17 ml. Dashed curve – $<m_x>$ of central 5-ml part of Fe$_{37}$Cr$_{63}$ layer. (d) $d_f$ dependence of effective Curie temperature, $T_C$. Dashed curve – $T_C$ of central part.

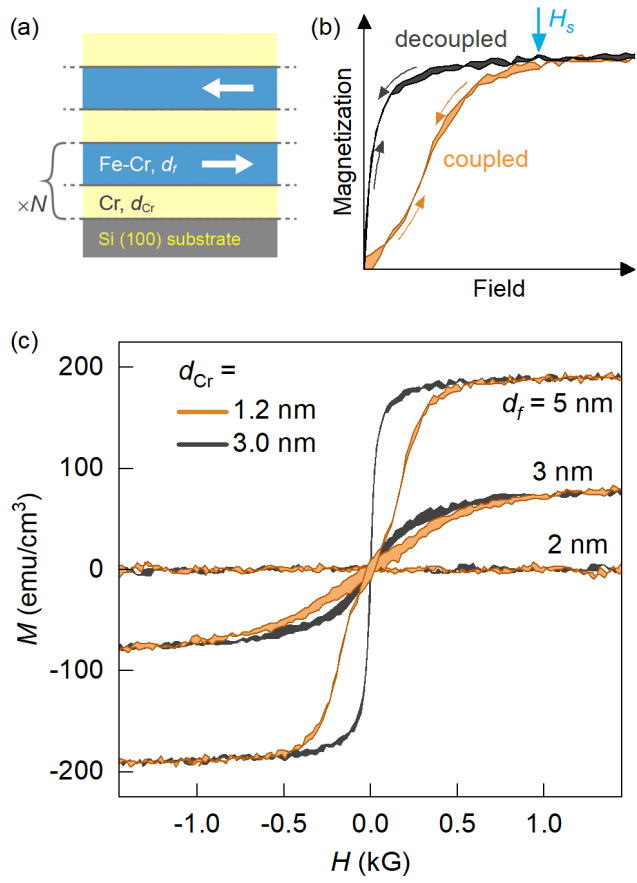

Fig.1.

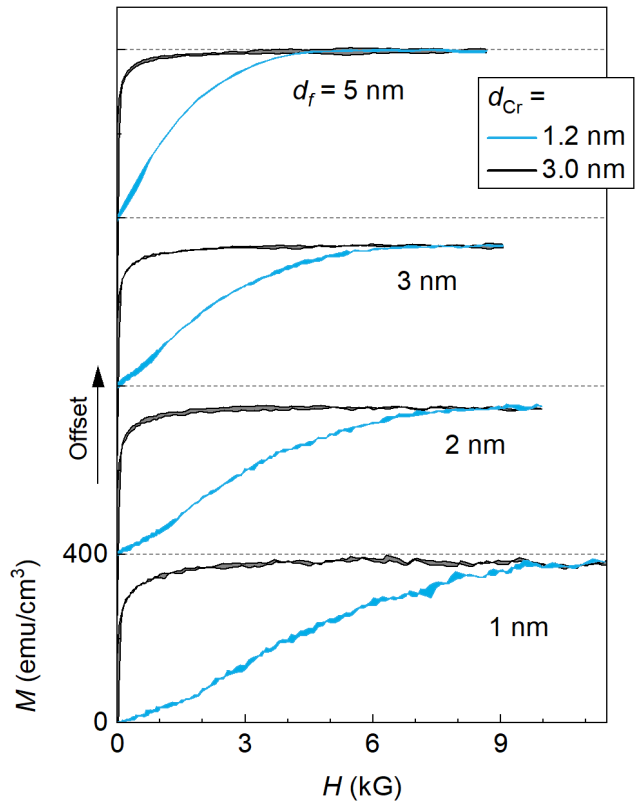

Fig. 2.

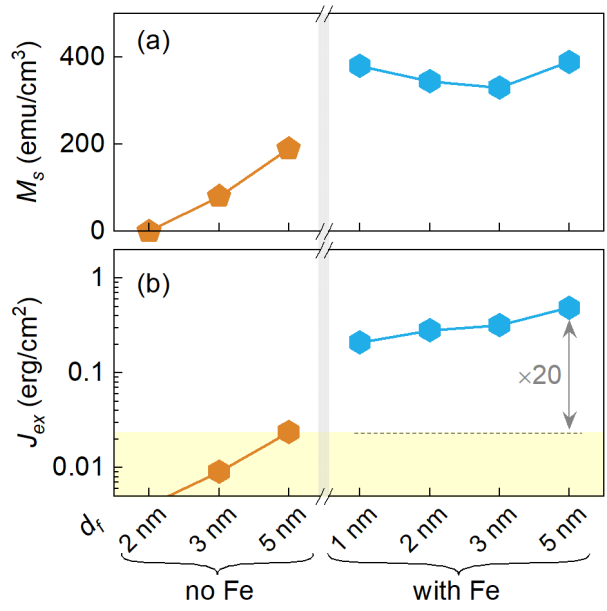

Fig. 3.

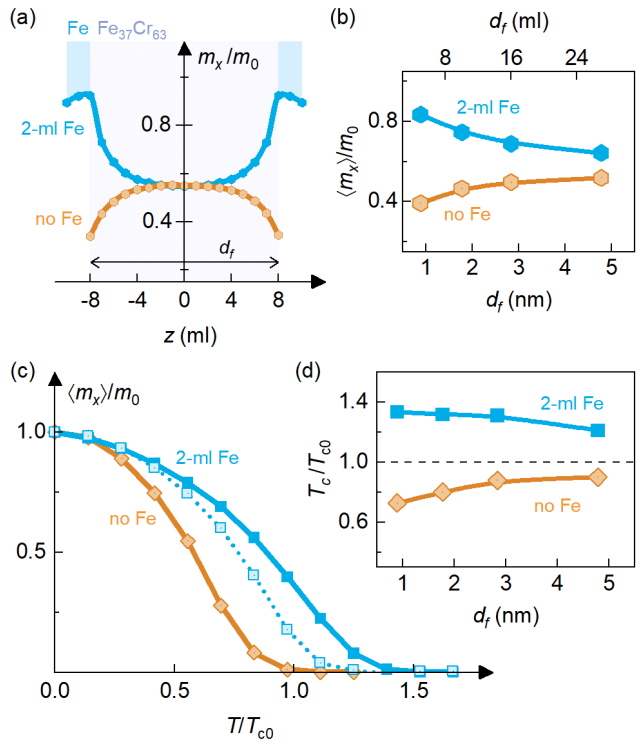

Fig. 4.